\documentclass[useAMS,usenatbib]{mn2e}
\usepackage[english]{babel}
\usepackage{aas_macros}
\usepackage{natbib}
\usepackage{amsmath}
\usepackage{graphicx}
\usepackage{hyperref}

\newcommand{\lcdm}{\mathrm{\Lambda CDM}}
\newcommand{\densm}{\Omega_{\mathrm{m}}}
\newcommand{\densl}{\Omega_{\mathrm{\Lambda}}}
\newcommand{\densk}{\Omega_{\mathrm{k}}}
\newcommand{\der}{\mathrm{d}}
\DeclareMathOperator{\sinn}{sinn}

\title[Computing the Luminosity Distance]{Numerical Strategies of Computing the 
Luminosity Distance}

\author[Liu et al.]{De-Zi Liu$^{1}$, Cong Ma$^{1}$, Tong-Jie 
Zhang\thanks{E-mail: tjzhang@bnu.edu.cn}$^{1,2}$ and Zhi-Liang Yang$^{1}$\\
$^{1}$Department of Astronomy, Beijing Normal University, Beijing 100875, 
P.~R.~China\\
$^{2}$Center for High Energy Physics, Peking University, Beijing 100871, 
P.~R.~China}

\begin{document}

\date{}

\maketitle

\label{firstpage}

\begin{abstract}
    We propose two efficient numerical methods of evaluating the luminosity 
    distance in the spatially flat $\lcdm$ universe.  The first method is based 
    on the Carlson symmetric form of elliptic integrals, which is highly 
    accurate and can replace numerical quadratures.  The second method, using a 
    modified version of Hermite interpolation, is less accurate but involves 
    only basic numerical operations and can be easily implemented.  We compare 
    our methods with other numerical approximation schemes and explore their 
    respective features and limitations.  Possible extensions of these methods 
    to other cosmological models are also discussed.
\end{abstract}


\section{Introduction}

The computation of cosmological distances naturally arises in the study of 
cosmology, for example the luminosity distance $d_L$ in the analysis of type Ia 
supernova (SNIa) data and the angular diameter distance $d_A$ in the study of 
gravitational lensing.  These distances depend on the underlying cosmological 
model and their parameters.  Therefore they are useful as cosmological tests.  
As a result, accurate and efficient numerical algorithms of evaluating these 
distances become a necessity for the practitioners of cosmological research.

The analytical form of the cosmological distances can be derived from the 
solution of the Friedmann equation, an ordinary differential equation involving 
the scale factor $a$ as a function of cosmic time $t$.  Therefore, formulae for 
the distances usually involve an integral over the expansion history expressed 
in terms of the redshift $z$ and cosmological parameters.  In general, the 
integrations can be evaluated numerically by quadrature algorithms.  However, 
numerical quadratures tend to be computationally heavy when high accuracy is 
desired.

\defcitealias{1999ApJS..120...49P}{Pen99}
\defcitealias{2010MNRAS.406..548W}{WU10}
In the presence of this performance issue, it is advantageous to develop 
algorithms that are restricted to specific cosmological models but are 
otherwise more efficient than general-purpose quadratures.  For the spatially 
flat $\lcdm$ model, efficient algorithms for the luminosity distnace have been 
proposed by \citet[henceforward Pen99]{1999ApJS..120...49P} and 
\citet[henceforward WU10]{2010MNRAS.406..548W}.

In this paper we propose two different numerical methods for the luminosity 
distance also in the context of the spatially flat $\lcdm$ universe.  The 
methods are presented in Sections 2 and 3 respectively.  In Section 4, the 
performances of these methods are discussed.  Finally in Section 5 we discuss 
some possible extensions to the methods presented in this paper.  Throughout 
the paper we will focus on the luminosity distance only, but the results can be 
trivially extended to compute the angular diameter distance $d_A = d_L / (1 + 
z)^2$.

\section{Method I: Evaluation Using Carlson Symmetric Forms}

The luminosity distance $d_L$ in the spatially flat $\lcdm$ universe is given 
by
\begin{eqnarray}
    \label{eq:dlflat}
    d_L(z) = \frac{c(1 + z)}{H_0} \int_{0}^{z} \frac{\der t}{\sqrt{\densm (1 + 
    t)^3 + \densl}},
\end{eqnarray}
where $\densm$ and $\densl$ are the energy densities corresponding to the 
matter and cosmological constant respectively: $\densm + \densl = 1$.  
Following the notation in \citetalias{1999ApJS..120...49P} we introduce the 
parameter $s = \sqrt[3]{(1 - \densm) / \densm}$ and the change-of-variable $u = 
1 / t$, and re-write equation (\ref{eq:dlflat}) as
\begin{eqnarray}
    \label{eq:dlflat-s}
    \frac{d_L}{c / H_0} = \frac{1 + z}{\sqrt{s\densm}} \left[T(s) - 
    T\left(\frac{s}{1+z}\right)\right],
\end{eqnarray}
where
\begin{eqnarray}
    \label{eq:tint}
    T(x) = \int_{0}^{x} \frac{\der u}{\sqrt{u^4 + u}}.
\end{eqnarray}

The integral in equations (\ref{eq:dlflat}) and (\ref{eq:tint}) are special 
cases of elliptic integrals.  All elliptic integrals can be reduced to several 
basic forms, the best known of which are probably the three kinds of Legendre 
elliptic integrals \citep[Chapter 22]{wwcourse}, with reduction theorems and 
examples presented in \citep[Chapter 17]{asbook}.  In our case it is clearer to 
express this integral by one of the Carlson symmetric forms $R_F(x_1, x_2, 
x_3)$, which is defined as
\begin{eqnarray}
    \label{eq:defrf}
    R_F(x_1, x_2, x_3) = \frac{1}{2} \int_{0}^{+\infty}\!\!\!\frac{\der 
    t}{\sqrt{(t + x_1) (t + x_2) (t + x_3)}}.
\end{eqnarray}
Using the reduction theorems\footnote{See \citep[Chapter 19]{nisthmf}, 
available online at \mbox{\url{http://dlmf.nist.gov/19.29}}}, it is 
straightforward to verify that
\begin{eqnarray}
    \label{eq:redtocarlson-d}
    T(x) = 4 R_F(m, m + 3 + 2 \sqrt{3}, m + 3 - 2 \sqrt{3}),
\end{eqnarray}
where
\begin{eqnarray}
    m(x) = \frac{2 \sqrt{x^2 - x + 1}}{x} + \frac{2}{x} - 1. \nonumber
\end{eqnarray}

It has been known that the Carlson forms can be computed numerically with high 
accuracy.  \citet{Carlson79} showed that the computation of $R_F$ can be 
accomplished iteratively with the error decreasing by a factor of $4^6$ after 
each iteration, therefore achieving fast convergence.  Further analysis of the 
algorithms for $R_F$ and other elliptic integrals can be found in 
\citep{Carlson94}, and computer implementation details have been discussed in 
\citep{Algorithm577} and \citep[Chapter 6]{NR3}.

\section{Method II: Approximation by a Modified Hermite 
Interpolation}

The method presented in Section 2 uses an iterative approach to the computation 
of $R_F$.  However, there are situations where a closed, approximate formula 
for the integral in equation (\ref{eq:tint}) is desired.  In 
\citetalias{1999ApJS..120...49P} an approximation was obtained using polynomial 
fit for $T(x)$.  In \citetalias{2010MNRAS.406..548W} another method with higher 
accuracy was proposed.  In this section we show how a modified version of 
Hermite interpolation can lead to a class of approximations similar to that in 
\citetalias{1999ApJS..120...49P}.

We intend to approximate equation (\ref{eq:tint}) using only elementary 
operations, such as polynomial evaluation and $n$th root where $n$ is a small 
integer.  We note that the behavior of $T(x)$ has several deficiencies.  First, 
the derivative of $T(x)$ becomes singular as $x \to 0^{+}$.  Second, the domain 
of $T(x)$ extends to infinity.  Either one is detrimental to the approximation 
using polynomials.  However, they can be removed by certain 
change-of-variables.  For example, we can introduce a new function
\begin{eqnarray}
    \label{eq:xidef}
    \xi(x) = T^2(\frac{1}{x} - 1)
\end{eqnarray}
that has smooth derivatives within the interval $0 < x < 1$ and can be extended 
to the cases of $x \to 0^{+}$ and $x \to 1^{-}$.  The limiting behaviors of 
$\xi(x)$ are shown below:
\begin{eqnarray}
    \label{eq:xiendpoints}
    \xi(0^{+}) = A^2, &\quad& \xi(1^{-}) = 0, \nonumber \\
    \xi'(0^{+}) = -2A, &\quad& \xi'(1^{-}) = -4,
\end{eqnarray}
where $A = T(+\infty) = 2.80436\cdots$ is a numerical constant\footnote{We note 
in passing that the constant $X$ in \citetalias{1999ApJS..120...49P}, equation 
(5) is identical to $1/A$.  A typo was made therein, which should have been $X 
\equiv [\int_{0}^{\infty}\!\!\der u/\sqrt{u^4 + u}]^{-1}$.}.

Using the end-point conditions in equation (\ref{eq:xiendpoints}) one can 
construct a 3rd-order polynomial, which is a linear combination of the four 
Hermite basis splines in $[0, 1]$, as a crude approximation with $\sim$20\% 
relative error.  This linear combination is unique, allowing no further 
improvements.  However, we note that for realistic values of $\densm$ it is not 
necessary to approximate $\xi(x)$ in the entire interval $[0, 1]$, because the 
subinterval $[0, \frac{1}{s + 1})$ corresponds to the scenario of $z < 0$, 
i.e.~``the future''.  Therefore, we can introduce a free parameter $x_{\ast}$ 
as the alternative lower end-point, and only perform the approximation in the 
subinterval $[x_{\ast}, 1]$, if a constraint is put on $\densm$ (or 
equivalently, $s$).

To accommodate further refinements, a correction term $w(x)$ can be added to 
the Hermite approximation.  We require the value and first derivative of $w(x)$ 
to vanish at either end-point, so that it can be added to the Hermite 
approximation without altering the coefficients on the basis splines.  One 
choice of $w(x)$ is made possible by a family of functions
\begin{eqnarray}
    \label{eq:corr}
    w(x) = x^2 (1-x)^2 (ax + b + 2a)
\end{eqnarray}
where $a$ and $b$ are adjustable parameters accounting for the deviation of the 
Hermite approximation from the true function.  Other choices are possible, but 
we will begin with the simple case of equation (\ref{eq:corr}).

By construction, the approximation described above has the property that the 
approximating function coincides with the true function at the end-points, 
$x_{\ast}$ and $1$, up to the first derivative.  But we note that the goal is 
to approximate equation (\ref{eq:dlflat-s}) rather than equation 
(\ref{eq:tint}).  This suggests that the implicit requirement of the 
coincidence of function values at end-points could be unnecessarily strong.  
Alternatively, we may refrain from requiring the approximating function values 
to match the true ones.  Instead, we only require the matching of first 
derivatives at $x = x_{\ast}$, and leave the end-point value at another free 
parameter.  To summarize, we now have four free parameters that can be tuned: 
$x_{\ast}$, $a$ and $b$, and the function value $v_0$ at $x = x_{\ast}$.  The 
approximation to equation (\ref{eq:xidef}) can be expressed as
\begin{eqnarray}
    \label{eq:happrox}
    \tilde{\xi}(y) = v_0 H^{(0)}_{0}(y) + l \left[d_0 H^{(1)}_{0}(y) -4 
    H^{(1)}_{1}(y)\right] + w(y),
\end{eqnarray}
where
\begin{eqnarray}
    l = 1 - x_{\ast}, \quad y = \frac{x - x_{\ast}}{l}, \quad d_0 = 
    \xi'(x_{\ast}), \nonumber
\end{eqnarray}
and $H_{i}^{(j)}$ are the Hermite basis splines,
\begin{eqnarray}
    H^{(0)}_{0}(y) &=& 2 y^3 - 3 y^2 + 1, \nonumber \\
    H^{(0)}_{1}(y) &=& -2 y^3 + 3 y^2, \nonumber \\
    H^{(1)}_{0}(y) &=& y^3 - 2 y^2 + y, \nonumber \\
    H^{(1)}_{1}(y) &=& y^3 - y^2. \nonumber
\end{eqnarray}

Following the approach in \citetalias{1999ApJS..120...49P}, we choose the 
objective function as the maximum relative error in $d_L$ using the 
approximation ([Eq.~\ref{eq:happrox}]), with the restriction $0.2 \le \densm 
\le 1$.  Minimizing the objective function over the parameters, we obtain the 
best-fit $x_{\ast} = 0.40176$, $a = 1.62053$, $b = -6.34985$, and $v_0 = 
4.64111$.  Substituting the numerical values into equation (\ref{eq:happrox}), 
we therefore construct an approximation polynomial
\begin{eqnarray}
    \label{eq:happrox-n}
    \tilde{\xi}(y) = 1.62053 y^5 - 6.34985 y^4 + 8.41443 y^3 \nonumber \\
    - 2.01328 y^2 - 6.31293 y + 4.64111.
\end{eqnarray}
Equation (\ref{eq:happrox-n}) is the main result of this section.  With the 
parameters determined, the approximation to $d_L$ can be computed using this 
formula with equations (\ref{eq:dlflat-s}) and (\ref{eq:xidef}).

\section{Performance of the Methods}

In this section we proceed to assess the performance of the methods in Sections 
2 and 3.  The assessment is mainly done in terms of the accuracy and 
efficiency.

\subsection{Accuracy}

The first method can be used to yield highly accurate numerical approximation 
of $d_L$ for vast ranges of $z$ and the parameter $s$ if we adopt the algorithm 
for $R_F$ by \citet{Carlson79, Carlson94}.  Unlike the methods based on the 
evaluation of a closed approximation formula, the desired cutoff error can be 
prescribed to determine when the iterative computation of $R_F$ terminates.  In 
practice, we found that the prescription of relative error $\sim$10$^{-16}$ can 
be achieved without suffering significant loss in the computation speed.

\begin{figure}
    \includegraphics[width=.5\textwidth]{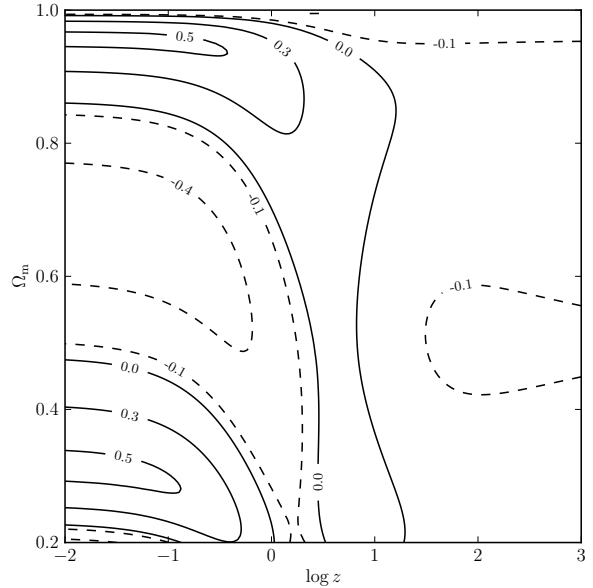}
    \caption{Contour plot for the distribution of the relative error in $d_L$ 
    using the approximation method in Section 3.  Positive and negative values 
    of the error are plotted in solid and dashed lines respectively.  The 
    ``peaks'' and ``pits'' in the left region of this figure ($z < 0.1$) 
    dominate
    the global error.}
    \label{fig:globalerr}
\end{figure}

For the second method, we plot the distribution of the relative error of $d_L$ 
in Figure \ref{fig:globalerr}.  As can be seen from the figure, the second 
method remains an approximation at best.  Under our choice of fitting 
parameters and range of $\densm$, the relative error in $d_L$ is $\sim$0.5\%.  
The major source of this error is contributed by $z < 0.1$.  For $0.1 < z < 10$ 
our method is comparable with that of \citetalias{1999ApJS..120...49P}, and 
ours slightly outperforms it when $z$ is larger.

\subsection{Efficiency}

Theoretically, the best-, worst-, and average-case temporal efficiencies for 
each method can be calculated or estimated by tracking every operation taken 
during the course of the computing.  However, such a thorough analysis is 
beyond the scope of this paper.  Instead, we empirically compare the running 
time of the computer programs using the two methods with those of 
\citetalias{1999ApJS..120...49P} and \citetalias{2010MNRAS.406..548W} under a 
controlled environment.

\begin{figure}
    \includegraphics[width=.5\textwidth]{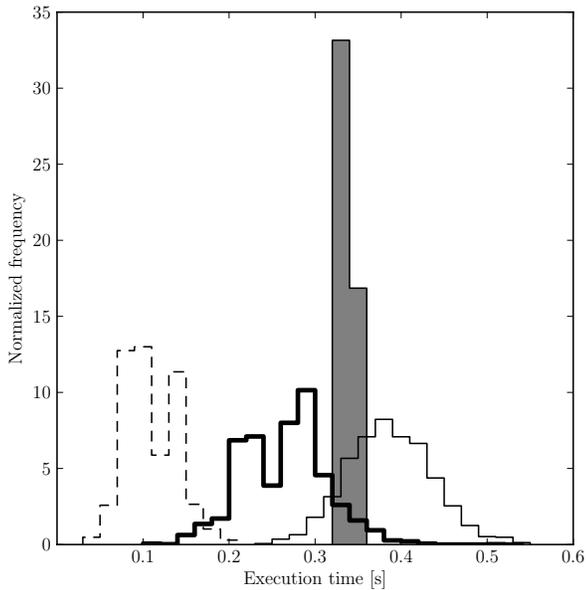}
    \caption{Histograms of the benchmark results for the four methods.  Dotted, 
    thick solid, shaded, and thin solid histograms represent the execution 
    timings of codes implementing \citetalias{1999ApJS..120...49P}, our Method 
    I, Method II, and \citetalias{2010MNRAS.406..548W} respectively.  Each 
    histogram is normalized so that the total probability (the area enclosed 
    under the boundaries) is unity.}
    \label{fig:perf}
\end{figure}

In Figure \ref{fig:perf} we display the benchmark results of our methods 
compared with that of \citetalias{1999ApJS..120...49P} and 
\citetalias{2010MNRAS.406..548W}.  To simulate a ``real-world'' application of 
these methods, we creates a sample of SNIa redshifts using the 
Supernova/Acceleration Probe \citep{2004astro.ph..5232S} fiducial redshift 
distribution containing $1998$ redshift points distributed within $0.1 \le z 
\le 1.7$ \citep[see][Table 1]{2006MNRAS.366.1081S}.  Our sample satisfies the 
same distribution to the SNAP fiducial, but is 16 times as dense, i.e.~with 
$31968$ points in total.  We have made custom implementations of the methods 
from \citetalias{1999ApJS..120...49P}, \citetalias{2010MNRAS.406..548W}, and 
our Method II in the C programming language, and uses the GNU Scientific 
Library (GSL\footnote{\mbox{\url{http://www.gnu.org/software/gsl/}}}) 
implementation of the $R_F$ algorithm in \citep{Algorithm577} for Method I.  In 
our benchmark routine, the computing of $d_L$ values from our redshift sample 
is performed for $\densm = 0.3, 0.5, 0.7, \mathrm{and } 0.9$ respectively, with 
each pass through the $z$ sample repeated for 25 times (that sums up to a total 
of $3.2\times10^6$ evaluations of $d_L$).  The benchmark itself is repeated for 
2400 times.

To interpret Figure \ref{fig:perf}, we make two remarks.  First, the execution 
time results were collected from the output of the {\tt gprof} 
profiler\footnote{\mbox{\url{http://www.gnu.org/software/binutils/}}} and does 
not reflect the absolute time spent.  It is only meaningful as a relative 
measure useful for comparing the speed of the codes with each other.  Second, 
the results are dependent on our particular implementations as well as the 
computing environment.  This is evident if our Figure \ref{fig:perf} is 
compared with Figure 4 in \citetalias{2010MNRAS.406..548W} that shows a 
reversed result for the speeds of the two methods in 
\citetalias{1999ApJS..120...49P} and \citetalias{2010MNRAS.406..548W}.

\section{Discussion}

As Figure \ref{fig:perf} suggests, both methods proposed in this paper is 
slower than the \citetalias{1999ApJS..120...49P} method.  However, Method I is 
a very reasonable trade-off between an enormous gain in accuracy and small loss 
of efficiency.  With Method I one does not need to resort to the numerical 
quadrature for the same level of accuracy.

Method I can be extended to cover the $\lcdm$ model with a curvature term 
$\densk$, because in that case the equivalent of equation (\ref{eq:dlflat}) 
assumes the form
\begin{eqnarray}
    \label{eq:dlnonflat}
    \frac{d_L}{c/H_0} &=& \frac{1 + z}{\sqrt{\left|\densk\right|}} \sinn \left[ 
    \sqrt{\left|\densk\right|} \int_{0}^{z}\frac{\der t}{E(t; \densm, \densk)} 
    \right]
\end{eqnarray}
where
\begin{eqnarray}
    E(z; \densm, \densk) = \sqrt{\densm (1+z)^3 + \densk (1+z)^2 + \densl} 
    \nonumber
\end{eqnarray}
is the expansion rate ($\densl = 1 - \densm - \densk$), and
\begin{eqnarray}
    \sinn(x) =
    \begin{cases}
	\sin(x), & \densk < 0;	\\
	x, & \densk = 0;	\\
	\sinh(x), & \densk > 0.
    \end{cases} \nonumber
\end{eqnarray}
The integral in equation (\ref{eq:dlnonflat}) is also an elliptic integral and 
can be reduced to $R_F$ accordingly.  This is potentially useful for the 
analysis of future SNIa data, because it has been suggested that the spacetime 
curvature should not be ignored in the probe of dark energy using luminosity 
distance data \citep{2007JCAP...08..011C,2008IJTP...47.2464O}.

In contrast, Method II may not be as promising, because in its current form the 
accuracy does not outperform that of \citetalias{1999ApJS..120...49P}.  
However, the idea behind the method may be useful when extending to alternative 
cosmological models (for example, dynamical dark energy) which may not be 
reduced to the applicable scenarios of Method I.  In the description of this 
method we have left some arbitrariness unjustified, notably the particular 
choice of the singularity-removing transformation (Eq.~[\ref{eq:xidef}]), the 
parameterization of the correction term (Eq.~[\ref{eq:corr}]), and the very 
choice of Hermite basis splines.  Alternative choices of them may be adopted to 
generate better approximations, for instance, the use of low-order 
Hermite-Birkhoff interpolation\footnote{See \citep{ahb68} for the general 
theory, and Problem 8.9 of \citep{kressbook} for a low-order example.} to 
selectively choose the point $x \in [0, 1]$ near which the derivative 
information of the true function is to be best preserved.  Moreover, our Method 
II uses only elementary numerical operations, while in 
\citetalias{2010MNRAS.406..548W} the numerical logarithm is extensively used.

\section*{Acknowledgements}

De-Zi Liu would like to thank Fang-Fang Zhu and Mao-Sheng Xiang for their kind 
help. This work was supported by the National Science Foundation of China 
(Grants No.~10473002), the Ministry of Science and Technology National Basic 
Science program (project 973) under grant No.~2009CB24901, the Fundamental 
Research Funds for the Central Universities.

\bibliographystyle{mn2e}
\bibliography{ms}

\begin{thebibliography}{}

\bibitem[\protect\citeauthoryear{{Abramowitz} \& {Stegun}}{{Abramowitz} \&
  {Stegun}}{1972}]{asbook}
{Abramowitz} M.,  {Stegun} I.~A.,  eds, 1972, Handbook of Mathematical
  Functions.
U.S.~Gov.~Printing Office, Washington, D.C.

\bibitem[\protect\citeauthoryear{{Carlson}}{{Carlson}}{1979}]{Carlson79}
{Carlson} B.~C.,  1979, Numerische Mathematik, 33, 1

\bibitem[\protect\citeauthoryear{{Carlson}}{{Carlson}}{1994}]{Carlson94}
{Carlson} B.~C.,  1994, preprint, arXiv:math/9409227

\bibitem[\protect\citeauthoryear{{Carlson} \& {Notis}}{{Carlson} \&
  {Notis}}{1981}]{Algorithm577}
{Carlson} B.~C.,  {Notis} E.~M.,  1981, {ACM} Transactions on Mathematical
  Software, 7, 398

\bibitem[\protect\citeauthoryear{{Clarkson}, {Cort{\^e}s} \&
  {Bassett}}{{Clarkson} et~al.}{2007}]{2007JCAP...08..011C}
{Clarkson} C.,  {Cort{\^e}s} M.,    {Bassett} B.,  2007, J. Cosmol. Astropart.
  Phys., 8, 11

\bibitem[\protect\citeauthoryear{{Kress}}{{Kress}}{1998}]{kressbook}
{Kress} R.,  1998, Numerical Analysis.
Springer-Verlag, New York

\bibitem[\protect\citeauthoryear{{Olver}, {Lozier}, {Boisvert} \&
  {Clark}}{{Olver} et~al.}{2010}]{nisthmf}
{Olver} F.~W.~J.,  {Lozier} D.~W.,  {Boisvert} R.~F.,    {Clark} W.~C.,  eds,
  2010, {NIST} Handbook of Mathematical Functions.
Cambridge Univ.~Press, New York

\bibitem[\protect\citeauthoryear{{{\"O}zta{\c s}}, {Smith} \&
  {Paul}}{{{\"O}zta{\c s}} et~al.}{2008}]{2008IJTP...47.2464O}
{{\"O}zta{\c s}} A.~M.,  {Smith} M.~L.,    {Paul} J.,  2008, International
  Journal of Theoretical Physics, 47, 2464

\bibitem[\protect\citeauthoryear{{Pen}}{{Pen}}{1999}]{1999ApJS..120...49P}
{Pen} U.-L.,  1999, \apjs, 120, 49

\bibitem[\protect\citeauthoryear{{Press}, {Teukolsky}, {Vetterling} \&
  {Flannery}}{{Press} et~al.}{2007}]{NR3}
{Press} W.~H.,  {Teukolsky} S.~A.,  {Vetterling} W.~T.,    {Flannery} B.~P.,
  2007, Numerical Recipes: The Art of Scientific Computing, 3 edn.
Cambridge Univ.~Press, Cambridge, UK

\bibitem[\protect\citeauthoryear{{Shafieloo}, {Alam}, {Sahni} \&
  {Starobinsky}}{{Shafieloo} et~al.}{2006}]{2006MNRAS.366.1081S}
{Shafieloo} A.,  {Alam} U.,  {Sahni} V.,    {Starobinsky} A.~A.,  2006, \mnras,
  366, 1081

\bibitem[\protect\citeauthoryear{{Sharma} \& {Prasad}}{{Sharma} \&
  {Prasad}}{1968}]{ahb68}
{Sharma} A.,  {Prasad} J.,  1968, {SIAM} Journal on Numerical Analysis, 5, 864

\bibitem[\protect\citeauthoryear{{{SNAP}~{Collaboration}}}{{{SNAP}~{Collaborat%
ion}}}{2004}]{2004astro.ph..5232S}
{{SNAP}~{Collaboration}} 2004, preprint, arXiv:astro-ph/0405232

\bibitem[\protect\citeauthoryear{{Whittaker} \& {Watson}}{{Whittaker} \&
  {Watson}}{1969}]{wwcourse}
{Whittaker} E.~T.,  {Watson} G.~N.,  1969, A Course of Modern Analysis, 4 edn.
Cambridge Univ.~Press, New York

\bibitem[\protect\citeauthoryear{{Wickramasinghe} \&
  {Ukwatta}}{{Wickramasinghe} \& {Ukwatta}}{2010}]{2010MNRAS.406..548W}
{Wickramasinghe} T.,  {Ukwatta} T.~N.,  2010, \mnras, 406, 548

\end{thebibliography}

\label{lastpage}

\end{document}